\documentclass[12pt]{article}
\catcode`\@=11
\@addtoreset{equation}{section}

\global\arraycolsep=2pt
\usepackage{mathrsfs,amsbsy,amssymb,latexsym,amsfonts,amsmath,cite,braket}
\usepackage{graphicx,color,setspace}

\newcommand\no{\nonumber}

\newcommand{\dd}{{\rm d}}

\newcommand{\bomega}{\bar{\omega}}

\newcommand{\bnabla}{\bar{\nabla}}
\newcommand{\bg}{\bar{g}}
\newcommand{\bphi}{\bar{\phi}}
\newcommand{\bpsi}{\bar{\psi}}

\allowdisplaybreaks

\begin{document}

\begin{flushright}
\parbox{4cm}
{KUNS-2810
}
\end{flushright}

\vspace*{1cm}

\begin{center}
{\Large \bf  $T\bar{T}$-deformation and Liouville gravity}
\vspace*{1.5cm}\\
{\large  
Suguru Okumura\footnote{E-mail:~s.okumura@gauge.scphys.kyoto-u.ac.jp}
and Kentaroh Yoshida\footnote{E-mail:~kyoshida@gauge.scphys.kyoto-u.ac.jp}} 
\end{center}

\vspace*{0.4cm}

\begin{center}
{\it Department of Physics, Kyoto University, \\ 
Kitashirakawa Oiwake-cho, Kyoto 606-8502, Japan}  
\end{center}

\vspace{0.3cm}

\begin{abstract}
We consider a gravitational perturbation of the Jackiw-Teitelboim (JT) gravity  
with an arbitrary dilaton potential and study the condition under which 
the quadratic action can be seen as a $T\bar{T}$-deformation of 
the matter action. As a special case, the flat-space JT gravity discussed by Dubovsky et al 
[arXiv:1706.06604 ] is included. Another interesting example is a hyperbolic dilaton potential.  
This case is equivalent to a classical Liouville gravity with a negative cosmological constant and 
then a finite $T\bar{T}$-deformation of the matter action is realized as a gravitational 
perturbation on AdS$_2$.  
\end{abstract}

\setcounter{footnote}{0}
\setcounter{page}{0}
\thispagestyle{empty}

\newpage

\tableofcontents

\section{Introduction}

An intriguing subject is to understand $T\bar{T}$-deformations of 2D quantum field theory (QFT). 
An infinitesimal $T\bar{T}$-deformation is triggered by a composite operator 
$\alpha \det\left(T_{\mu\nu}^{[0]}\right)$\,, 
where $T_{\mu\nu}^{[0]}$ is the energy-momentum tensor of the original system 
and $\alpha$ is a constant parameter of dimension (length)$^2$\,. 
Hence this is an irrelevant perturbation of QFT. 
It has been elucidated in a seminal paper by Sasha Zamolodchikov \cite{Z} that 
this determinant operator is well-defined in two dimensions and the expectation values  
of the operator for the arbitrary (non-degenerate) energy eigenstates 
exhibit the factorization property under some basic assumptions. 
The finite-version of $T\bar{T}$-deformation 
is described by the following $T\bar{T}$-flow equation \cite{SZ,Tateo}:\footnote{The closed form of 
$T\bar{T}$-deformation is discussed in \cite{Bonelli}.} 
\begin{equation} 
\frac{{\rm d} \mathcal{L}^{[\alpha]}}{{\rm d} \alpha} = -\frac{1}{4}\det\left(T_{\mu\nu}^{[\alpha]}\right)\,. 
\end{equation}
Note here the energy-momentum tensor is for the deformed Lagrangian $\mathcal{L}^{[\alpha]}$\,. 
For a nice review on the $T\bar{T}$-deformation, see \cite{review}. 

\medskip 

Another interesting aspect of $T\bar{T}$-deformation is an intimate connection to 
a 2D dilaton gravity (which is often called the Jackiw-Teitelboim (JT) gravity 
\cite{Jackiw,Teitelboim}\footnote{For reviews on 2D dilaton gravity, see \cite{V,Nojiri}.})\,. 
In the work by Dubovsky et al \cite{Dubovsky}, it has been shown that a gravitational perturbation 
in the flat-space JT gravity can be seen as a finite $T\bar{T}$-deformation of the matter sector. 
Then, a generalization of the work \cite{Dubovsky} in flat space to AdS$_2$ and dS$_2$ 
has been discussed in \cite{IOSY}. However, the discussion in \cite{IOSY} 
is restricted to the conformal matter case and hence only the infinitesimal 
$T\bar{T}$-deformation has been discussed. 

\medskip 

In this paper, we will explain how to remove this conformal matter condition. The point is  
to replace the dilaton potential utilized in \cite{AP} with the hyperbolic one considered in \cite{KOY}. 
In particular, the hyperbolic dilaton potential model is equivalent to a classical Liouville gravity 
with a negative cosmological constant \cite{Frolov, regular}. Hence, 
in other words, a gravitational perturbation in the classical Liouville 
gravity\footnote{Our analysis here 
is at the classical level. Fort the relation between $T\bar{T}$-deformation and non-critical string 
at the quantum level, see \cite{Verlinde-NC,Tolley, Haruna}.}  
can be seen as a finite $T\bar{T}$-deformation of the matter action defined on AdS$_2$\,. 

\medskip 

This paper is organized as follows.
In section 2, we study a gravitational perturbation in the JT gravity with an arbitrary dilaton potential. 
The quadratic action can be regarded as an infinitesimal $T\bar{T}$-deformation of the matter action 
with the conformal matter condition. This section contains a brief review of the work \cite{IOSY}.  
In section 3, we introduce the classical Liouville gravity with a negative cosmological constant 
and show that its gravitational perturbation can be seen as a finite $T\bar{T}$-deformation 
without taking the conformal matter condition.  
Section 4 is devoted to conclusion and discussion.
Appendix A explains how to derive a gravitationally dressed S-matrix.
Appendix B introduces the general vacuum solution in the Liouville gravity 
with a cosmological constant.

\section{JT gravity with conformal matter}

In this section, we will consider a gravitational perturbation of the JT gravity with an arbitrary dilaton 
potential and derive the quadratic action. As a result, we can figure out the condition under which 
the gravitational perturbation can be seen as a $T{\bar T}$-deformation of the matter action. 
For example, for the constant dilaton potential, it can be seen as a finite $T\bar{T}$-deformation 
as shown in \cite{Dubovsky}. When the dilaton potential is given by a sum of the constant and a cosmological constant as utilized in \cite{AP}, the conformal matter condition is necessary \cite{IOSY}.  
That is, an infinitesimal $T\bar{T}$-deformation is realized on AdS$_2$ or dS$_2$\,.

\subsection{Our setup and notation}

The classical action of the JT gravity in the Lorentzian signature is given 
by the sum of the dilaton gravity action 
$S_{\rm dg}[g_{\mu\nu},\phi]$ ($g_{\mu\nu}$:\,metric, $\phi$:\,dilaton) 
and the matter action $S_{m}[\psi,g_{\mu\nu},\phi]$ ($\psi$:\,matter):
\begin{align}
S[g_{\mu\nu},\phi,\psi]&=S_{\rm dg}[g_{\mu\nu},\phi] 
+ S_{m}[\psi,g_{\mu\nu},\phi]\,, 
\label{total action}\\ 
\begin{split}
S_{\rm dg}[g_{\mu\nu},\phi] 
&=\frac{1}{2\kappa}\int\! \dd^2 x\,\sqrt{-g}\,\left[\,\phi\, R-U(\phi)\,\right]\,.  
\label{gravity action}
\end{split}
\end{align}
Here $x^{\mu}=(x^0,x^1)=(t,x)$ and $\kappa \equiv 8 \pi G_N$\,, where $G_{N}$ is 
2D Newton constant. The dilaton potential $U(\phi)$ is an arbitrary scalar function now. 
The matter action $S_{m}$ may include a non-trivial dilaton coupling in general. 

\medskip

The equations of motion of this system are given by 
\begin{align}
&R-U'(\phi) +\frac{2\kappa}{\sqrt{-g}} \frac{\delta S_{m}}{\delta \phi} =0\,,\label{eq:eom1}\\
&\frac{1}{2}g_{\mu\nu}U(\phi)
-\left(\nabla_{\mu}\nabla_{\nu}\phi-g_{\mu\nu}\nabla^2 \phi\right)
=\kappa\,T_{\mu\nu}\,,\label{eq:eom2}
\end{align}
and the one for the matter field.
Here we have used the identity $R_{\mu\nu}=\frac{1}{2}g_{\mu\nu}R$ for the second equation. 
The energy-momentum tensor $T_{\mu\nu}$ for the matter field $\psi$ is defined as 
\begin{align}
T_{\mu\nu} \equiv -\frac{2}{\sqrt{-g}}\frac{\delta S_m}{\delta g^{\mu\nu}}\,.
\end{align}
When the matter action depends on the dilaton, 
$T_{\mu\nu}$ also depends on the dilaton. 

\medskip

The trace of (\ref{eq:eom2}) is given by 
\begin{align}
&\nabla^2\phi+U\left(\phi\right)
=\kappa\,g^{\mu\nu}T_{\mu\nu}\equiv\kappa T\,.\label{eq:eom2trace}
\end{align}
Here $T$ is a trace of the energy-momentum tensor $T^{\mu\nu}$\,. 
By using (\ref{eq:eom2trace})\,, the Einstein equation (\ref{eq:eom2}) can be rewritten as 
\begin{align}
-\nabla_\mu \nabla_\nu \phi  - \frac{1}{2}g_{\mu\nu}\,U\left(\phi\right) &=\kappa\,\left(T_{\mu\nu}-g_{\mu\nu}T\right)\,.
\label{eq:eom2substract}
\end{align}

\subsubsection*{The vacuum solution}

For later convenience, let us discuss the vacuum solution (i.e., $T_{\mu\nu}=0$). 
When the matter action $S_{m}$ is turned off,  the equations of motion (\ref{eq:eom1}) 
and (\ref{eq:eom2}) reduced to 
\begin{align}
\bar{R}-U'\left(\bar{\phi}\right)&=0\,,\label{eq:veom1}\\
-\bar{\nabla}_\mu \bar{\nabla}_\nu \bar{\phi} + \bar{g}_{\mu\nu} \bar{\nabla}^2 \bar{\phi} + \frac{1}{2} \bar{g}_{\mu\nu}\,U\left(\bar{\phi}\right) &=0\,.\label{eq:veom2}
\end{align}
Here we have denoted the vacuum solution as $\bg_{\mu\nu}$ and $\bphi$, and the covariant derivative $\bnabla$ is defined with $\bg_{\mu\nu}$.
The trace of (\ref{eq:veom2}) is given by 
\begin{align}
&\bar{\nabla}^2 \bar{\phi}+U\left(\bar{\phi}\right)=0\,.\label{eq:veom-phitrace}
\end{align}
Since the dilaton potential $U(\phi)$ can be deleted from (\ref{eq:veom2}) by using  
(\ref{eq:veom-phitrace})\,, the resulting expression is  
\begin{align}
\bar{\nabla}_{\mu}\bar{\nabla}_{\nu}\bar{\phi}=\frac{1}{2}\bar{g}_{\mu\nu} \bar{\nabla}^2 \bar{\phi}\,.  
\label{eq:veom-phimunu}
\end{align}
Given the explicit form of $U(\phi)$\,, the vacuum solution is also determined.

\subsection{The quadratic action}

Next, we shall consider a gravitational perturbation around the vacuum solution, 
\begin{align}
g_{\mu\nu}=\bg_{\mu\nu}+h_{\mu\nu}\,, 
\qquad\phi=\bphi+\sigma\,,\qquad \psi = 0 + \psi\,, 
\label{eq:fluctuation}
\end{align}
where $h_{\mu\nu}$ and $\sigma$ are fluctuations of metric and dilaton, respectively, 
and $\psi$ itself is regarded as a fluctuation around the trivial background $\bar{\psi}=0$\,. 
In the following, the indices in the perturbations are raised, lowered, and contracted with the background metric $\bg_{\mu\nu}$, say 
$h^{\mu\nu} \equiv \bg^{\mu\rho}\bg^{\nu\sigma}h_{\rho\sigma}$\,. 

\medskip

Let us expand the classical action $S[g_{\mu\nu}, \phi, \psi]$ in (\ref{total action}) 
in terms of the fluctuations (\ref{eq:fluctuation}): 
\begin{align}
S[g_{\mu\nu}, \phi, \psi] &= S^{(0)}+S^{(1)}+S^{(2)}+\cdots\, \no \\ 
&= S^{(0)}_{\rm dg}[\bg_{\mu\nu},\bphi] + S^{(1)}_{\rm dg}[\bg_{\mu\nu},\bphi;h_{\mu\nu},\sigma] 
+ S^{(2)}_{\rm dg}[\bg_{\mu\nu},\bphi;h_{\mu\nu},\sigma]  \no \\ 
& ~~~+ S^{(1)}_{\rm m}[\bg_{\mu\nu};\psi] 
+ S^{(2)}_{\rm m}[\bg_{\mu\nu};\psi,h_{\mu\nu}] + \cdots \,,
\label{eq:action-expanded}
\end{align}
where the superscript of $S^{(n)}$ denotes the order of fluctuation. The zeroth order part 
$S^{(0)}_{\rm dg}$ is the classical value of $S_{\rm dg}$ with the vacuum configuration. 
The first order action $S^{(1)}_{\rm dg}$ should vanish since the vacuum solution satisfies 
the equations of motion with $\bpsi =0$\,. For the matter sector, $S_{\rm m}^{(1)}$ 
describes the matter field action on the classical background $\bar{g}_{\mu\nu}=0$\,. 

\medskip

By expanding \eqref{eq:eom1} and \eqref{eq:eom2substract}, 
the equations of motion for $h_{\mu\nu}$ and $\sigma$ can be obtained as
\begin{align}
&\bnabla^{\mu}\bnabla^{\nu}h_{\mu\nu}-\bnabla^2 h
-\frac{1}{2}U'(\bphi)\,h -U''(\bphi) \sigma+\frac{2\kappa}{\sqrt{-\bg}}\frac{\delta S_{m}^{(1)}}{\delta \phi}\left(\bar{\phi}\right)=0\,,\label{eq:eom1f}\,\\
&\bnabla_{\mu}\bnabla_{\nu}\sigma+\frac{1}{2}\bg_{\mu\nu}\,U'(\bphi)\,\sigma\no\\
&=-\kappa\,(T^{(0)}_{\mu\nu}-\bg_{\mu\nu}\,T^{(0)})-\frac{1}{2}U(\bphi)\,h_{\mu\nu}
+\frac{1}{2}\bnabla^\rho \bphi\left(\bnabla_{\mu}h_{\rho \nu}+\bnabla_{\nu} h_{\rho \mu}
-\bnabla_{\rho}h_{\mu\nu}\right)\,,\label{eq:eom2f}
\end{align}
where ${T^{(0)}}_{\mu\nu}$ is the energy-momentum tensor for the matter theory described by $S_{\rm m}^{(1)}$\,. 

\medskip

Suppose that $h_{\mu\nu}$ takes the following form \cite{IOSY}:  
\begin{align}
h_{\mu\nu}=-2\kappa(T^{(0)}_{\mu\nu}-\bg_{\mu\nu}\,T^{(0)})\, k \,, 
\label{ansatz}
\end{align}
where $k$ is a constant parameter with dimension (length)$^2$ while $\kappa$ is dimensionless. 
This is a covariant version of the one employed in the flat-space JT case \cite{Dubovsky}.  

\medskip

As a result, the quadratic action $S^{(2)}$ is simplified as
\begin{align}
S^{(2)}
&=\int\! \dd^2 x\,
\sqrt{-\bar{g}}\,\biggl[\frac{1}{4\kappa}U''\left(\bar{\phi}\right) \sigma^2
-\kappa\left(k-\frac{k^2}{4}U\left(\bar{\phi}\right)\right)\left(\,T^{(0)}_{\mu\nu}{T^{(0)}}^{\mu\nu}-{T^{(0)}}^2\right)\,\biggr]\label{quadratic action}\,.
\end{align}
The second term is proportional to the $T\bar{T}$ operator, though the coefficient depends 
on the background dilaton $\bar{\phi}$ and in general has space-time coordinate dependence.

\medskip

Therefore, if $U''(\bphi)=0$ and the metric fluctuation $h_{\mu\nu}$ satisfies  
the ansatz (\ref{ansatz}), the quadratic action can be regarded as a $T\bar{T}$ deformation 
of the original matter action, up to the background dilaton dependence. 
However, we still need to check the existence of $\sigma$ as a solution 
to the equations of motion. 

\medskip 

In the following, let us see two examples, 1) the flat-space JT gravity and 
2) the Almheiri-Polchinski (AP) model

\subsubsection*{1) The flat-space JT gravity}

Let us first revisit the case of the flat-space JT gravity \cite{Dubovsky}. 
This case is realized by taking a constant dilaton potential
\begin{align}
U(\phi)=\Lambda \quad \mbox{~:~constant}\,.  
\end{align}
In the Cartesian coordinates, a vacuum solution is obtained as 
\begin{align}
\bar{g}_{\mu\nu} = \eta_{\mu\nu} = {\rm diag}(-1,+1)\,, \qquad 
\bphi=\frac{\Lambda}{4}(t^2-x^2)\,. 
\end{align}
Note here that the dilaton is non-trivial but the background metric is still 2D Minkowski 
spacetime. 

\medskip

The metric ansatz (\ref{ansatz}) satisfies (\ref{eq:eom1f}) only if the matter action 
$S_{m}^{(1)}$ does not depend on the dilaton
\begin{align}
\frac{\delta S_{m}^{(1)}}{\delta \phi}=0\,.
\end{align}
The solution for $\sigma$ is explicitly written down as a non-local solution \cite{IOSY}
\begin{equation}
\sigma(t,x) = a_1 + a_2 \,t + a_3\, x + \sigma_{\textrm{non-local}}(t,x)\,,
\end{equation}
where $a_{1,2,3}$ are arbitrary constants, and $\sigma_{\textrm{non-local}}$ 
is a non-local part of $\sigma$ given by 
\begin{align}
\sigma_{\textrm{non-local}}(t,x) &= \frac{\kappa}{2}
\Biggl[ \, k\Lambda\int^{x}_{0}\!\!\!\dd x'\,x'\,T^{(0)}_{tt}(t,x')+k\Lambda
\int^{t}_{0}\!\!\!\dd t'\,t'\,T^{(0)}_{xx}(t',x) \nonumber \\
&\qquad \quad 
+2\left(k\Lambda-1\right)\int^{t}_{0}\!\!\dd t'\int^{x}_{0}\!\!\dd x'\,T^{(0)}_{tx}(t',x') \nonumber \\
&\qquad \quad +\left(k\Lambda-2\right)
\left(\int^{t}_{t_1}\!\!\dd t'\int^{t'}_{t_2}\!\!\dd t''\,T^{(0)}_{xx}(t'',0) 
+\int^{x}_{x_1}\!\!\dd x'\int^{x'}_{x_2}\!\!\dd x''\,T^{(0)}_{tt}(0,x'')\right)
\Biggr]\,.
\label{sol_snl}
\end{align}
By using this expression of $\sigma$\,, one can introduce dynamical coordinates explicitly 
and evaluate the gravitationally dressed S-matrix \cite{Dubovsky}
(For the detail, see Appendix \ref{dressed S-matrix}).

\subsubsection*{2) The Almheiri-Polchinski model}

A bit non-trivial example satisfying the condition 
$U''(\bphi)=0$ is the Almheiri-Polchinski (AP) model \cite{AP} specified by 
the following dilaton potential: 
\begin{align}
U(\phi)=\Lambda-\frac{2}{L^2}\phi\,, 
\label{APpotential}
\end{align}
where $L$ is the AdS radius. In comparison to the flat-space JT case, a negative cosmological 
constant is additionally included. 

\medskip
 
In the conformal gauge, the metric is parametrized as 
\begin{align}
\dd^2 s=\bg_{\mu\nu}\dd x^{\mu} \dd x^{\nu}=-2 {\rm e}^{2\bar{\omega}}\dd x^{+} \dd x^{-}\label{conformal gauge}\,, 
\end{align}
where the light-cone coordinates are defined as 
\begin{eqnarray}
x^{\pm} \equiv \frac{1}{\sqrt{2}}( t \pm x )\,. 
\end{eqnarray}
The general vacuum solution is given by \cite{AP}
\begin{align}
{\rm e}^{2\bar{\omega}}
= \frac{2\,L^2}{(x^+-x^-)^2}\,,\qquad \bphi=\frac{\Lambda L^2}{2}+\frac{a+b(x^++x^-)+c\,x^+x^-}{x^+-x^-}\label{APgeneral}\,, 
\end{align}
where $a$, $b$ and $c$ are arbitrary constants. In the following discussion, 
we will consider only constant dilaton case with $a=b=c=0$ so as to drop off the coordinate 
dependence of the dilaton background. 

\medskip 

By imposing the metric ansatz (\ref{ansatz}), the equation of motion (\ref{eq:eom1f}) 
is rewritten as 
\begin{align}
\frac{{T^{(0)}}\,k}{L^2}+\frac{1}{\sqrt{-\bg}}
\frac{\delta S_{m}^{(1)}}{\delta \phi}\left(\bar{\phi}\right)=0\,.
\end{align}
A solution is to employ a conformal matter which does not couple to the dilaton \cite{IOSY}:  
\begin{align}
T^{(0)}=0,\,\qquad \frac{\delta S_{m}^{(1)}}{\delta \phi}=0\,.
\end{align}
Notably, this conformal matter condition is not a unique solution and 
there may remain another possibility to take a particular dilaton dependence 
of the matter action. We will discuss this issue in the next section.

\medskip 

Let us solve the equation of motion for $\sigma$ in (\ref{eq:eom2f})\,. 
Due to the conformal matter condition and the conservation law for 
the energy-momentum tensor, $T^{(0)}_{++}$ and $T^{(0)}_{--}$ 
are holomorphic and anti-holomorphic functions, respectively.
Each component of (\ref{eq:eom2f}) is evaluated as 
\begin{align}
{\rm e}^{2\bar{\omega}}\partial_+\left({\rm e}^{-2\bar{\omega}}\partial_+\sigma\right)=
& -\kappa\, T^{(0)}_{++}(x^+)\,,\no\\
{\rm e}^{2\bar{\omega}}\partial_-\left({\rm e}^{-2\bar{\omega}}\partial_-\sigma\right)=&
-\kappa\, T^{(0)}_{--}(x^-)\,,\no\\[-4pt]
\partial_+\partial_-\sigma+\frac{2\,\sigma}{(x^+-x^-)^2}=&\,0\,.\label{eq:AP}
\end{align}
A solution to the equations in (\ref{eq:AP}) is given by \cite{IOSY} 
\begin{align}
\sigma(x^+,x^-) \equiv \frac{I_0(x^+,x^-) + I^+(x^+,x^-) - I^-(x^+,x^-)}{x^+-x^-}\,.
\end{align}
Here $I_0$ is the sourceless solution, 
\[
I_0(x^+,x^-)\equiv A + B\,(x^++x^-) + C\, x^+x^-\,,  \qquad A,~B,~C \mbox{:~arbitrary real consts.}\,,  \label{sourceless part}
\] 
and $I^\pm(x^+,x^-)$ correspond to the non-local parts of dilaton: 
\begin{align}
I^\pm(x^+,x^-) \equiv &\, \kappa 
\int^{x^\pm}_{u^\pm}\!\!\dd s\, (s-x^+)(s-x^-)\,T^{(0)}_{\pm\pm}(s)\,, 
\label{API}
\end{align}
where $u^{\pm}$ are arbitrary constants.

\medskip

We should emphasize that in comparison to the flat-space JT gravity,  
the conformal matter condition $T^{(0)}=0$ is necessary here.
Because of the conformal matter condition, a general deformd system $\mathcal{L}^{(\alpha)}$ 
cannot be taken as the original matter action $S_{m}^{(1)}$\,. Hence a finite 
$T\bar{T}$-deformation cannot also be considered, though an infinitesimal $T\bar{T}$-deformation 
of a conformal field. This is a summary of the work \cite{IOSY}. 

\medskip 

Obviously, it is a significant issue to consider how to remove this conformal matter 
condition in the case of AdS$_2$\,.  In the next section, we will present another example 
supporting a non-conformal matter.

\section{Liouville gravity and $T\bar{T}$-deformation}

So far, we have considered at most a linear potential like (\ref{APpotential}) in order to 
solve the condition $U''(\phi)=0$\,. Note however that the condition we have to solve is that 
$U''(\bar{\phi}) =0$ and the argument is the background dilaton $\bar{\phi}$ rather than $\phi$\,. 
Hence it is enough to consider the behavior of the dilaton potential around the vacuum solution 
and it is possible to take account of more complicated dilaton potentials. 

\medskip

As such an example, we will consider a classical Liouville gravity with a negative cosmological constant.\footnote{The classical Liouville gravity can be derived from pure Einstein gravity in $2+\epsilon$ dimensions \cite{GJ}.}
Remarkably, the quadratic action is recast into a finite $T\bar{T}$-deformation of 
the original matter action (i.e., the conformal matter condition is not necessary).

\subsection{Classical Liouville gravity}

The classical action of the Liouville gravity with a negative cosmological constant is 
\begin{align}
S &= \frac{1}{2 \kappa} \int\! \dd ^2 x\, \sqrt{-G} \left( \phi R_{(\rm G)}-\frac{2\eta}{L^2}(\nabla_{(\rm G)} \phi)^2-\frac{1}{2\eta}{\rm e}^{2\eta\left(\Lambda-\frac{2}{L^2}\phi\right)}+\frac{1}{2\eta}\right)+S_{m}\,\left[\psi,\,G_{\mu\nu}\right]\,, 
\label{Liouville action}
\end{align}
where $R_{(G)}$ and $\nabla_{(G)}$ are the Ricci scalar and covariant derivative, respectively, 
defined with the metric $G_{\mu\nu}$\,. Then $\eta$ is a new constant parameter with dimension 
(length)$^2$\,. When $\eta$ is negative ($\eta<0$), the kinetic term of the dilaton has the wrong sign and the potential of (\ref{Liouville action}) is not bounded from below.
Hence, we take $\eta$ to be positive as a natural choice, 
\begin{align}
\eta>0\,.
\label{eta sign}
\end{align}

\medskip

In comparison to the action (\ref{gravity action})\,, 
the Liouville gravity action (\ref{Liouville action}) has the dilaton kinetic term. 
Hence, in order to employ the argument in Section 2, we have to remove the dilaton kinetic term 
by performing an appropriate Weyl transformation. 

\medskip

Let us consider the following Weyl transformation depending on the dilaton \cite{regular},
\begin{align}
G_{\mu\nu}={\rm e}^{-\eta\left(\Lambda-\frac{2}{L^2}\phi\right)} g_{\mu\nu}\,.
\end{align}
In the frame with $g_{\mu\nu}$\,, the kinetic term has been removed as follows: 
\begin{align}
S &= \frac{1}{2 \kappa} \int \dd ^2 x \sqrt{-g} \left( \phi R-\frac{1}{\eta}\text{sinh} \left[\eta
\left(\Lambda-\frac{2\phi}{L^2}\right)\right]\right)+S_{m}\,\left[\psi,\,
{\rm e}^{-\eta\left(\Lambda-\frac{2}{L^2}\phi\right)} g_{\mu\nu}\right]\,.\label{deformedAP action}
\end{align}
Thus the dilaton potential $U(\phi)$ in (\ref{gravity action}) is identified with the following 
hyperbolic potential:
\begin{equation}
U(\phi)=\frac{1}{\eta}\text{sinh} \left[\eta\left(\Lambda-\frac{2\phi}{L^2}\right)\right]
\label{deformedAP}\,.
\end{equation}
Note here that the matter action $S_{m}$ now depends on the dilaton explicitly 
through the Weyl factor of the metric $g_{\mu\nu}$\,. 

\medskip

Originally, this hyperbolic-type potential was introduced in \cite{KOY} so as to support 
Yang-Baxter deformations\cite{Klimcik, DMV, KMY} of AdS$_2$\,, 
where $\eta$ corresponds to the deformation parameter.
In the $\eta\rightarrow 0$ limit, the AP model (\ref{APpotential}) is reproduced.

\medskip

It is known that the AdS$_2$ metric with a constant dilaton is one of the vacuum solutions 
(For the general solution, see Appendix B).
In the conformal gauge (\ref{conformal gauge}), this solution is given by 
\begin{align}
{\rm e}^{2\bar{\omega}}
= \frac{2\,L^2}{(x^+-x^-)^2}\,,\qquad \bphi=\frac{\Lambda L^2}{2}\,. 
\end{align}
In the following, we will consider fluctuations around this vacuum solution.
For this constant dilaton background, one can show that 
\begin{align}
U(\bar{\phi})=0\,,\qquad U'(\bar{\phi})=-\frac{2}{L^2}\,,\qquad U''(\bar{\phi})=0\,. 
\end{align}
Thus this hyperbolic dilaton potential (\ref{deformedAP}) indeed satisfies the condition 
$U''(\bar{\phi})=0$\,.

\subsection{The quadratic action}

Let us then consider the quadratic action for the hyperbolic potential (\ref{deformedAP})\,.  
By supposing the ansatz (\ref{ansatz})\,, the equation in (\ref{eq:eom1f}) is simplified as
\begin{align}
\frac{2\kappa}{L^2}\,{T^{(0)}}\,k +\frac{2\kappa}{\sqrt{-\bg}}\frac{\delta S_{m}^{(1)}}{\delta \phi}\left(\bar{\phi}\right)=0\label{eq:eom1fsimple}\,.
\end{align}
The dilaton dependence in the matter action has been determined in (\ref{deformedAP action}), and the second term in (\ref{eq:eom1fsimple}) is replaced by the trace of the energy-momentum tensor as follows: 
\begin{align}
\frac{\delta S_{m}^{(1)}}{\delta \phi}(\bphi)&=-\frac{2\eta}{L^2}\,g^{\mu\nu}\frac{\delta S_{m}^{(1)}}{\delta g^{\mu\nu}}=\frac{\eta}{L^2}\sqrt{-\bg}\,{T^{(0)}}\,.
\end{align}
As a result, (\ref{eq:eom1fsimple}) reduces to a simple equation, 
\begin{align}
\frac{2\kappa}{L^2}\left(k+\eta\right)\,{T^{(0)}}=0\label{k condition}\,.
\end{align}
A possible solution is the conformal matter case ${T^{(0)}}=0$\,.
Then the matter action $S_{m}$ is invariant under the Weyl transformation  
and its dilaton dependence disappears. Hence, it is the same as the AP model case discussed 
in Section 2.

\medskip

Unless the matter action is conformal, $k$ is directly connected to $\eta$ like 
\begin{equation}
k=-\eta\,. \label{equiv}
\end{equation}
Originally, $k$ has been introduced as an arbitrary constant of the metric ansatz (\ref{ansatz})\,. 
In comparison to the flat-space JT case where $k$ is completely free,  
in the present case $k$ is determined completely by the initial set-up of the Liouville action. 

\medskip

With the condition (\ref{equiv}), the quadratic action (\ref{quadratic action}) leads to 
the form of $T\bar{T}$-deformation on the AdS$_2$ background,   
\begin{align}
S^{(2)}&=\kappa\,\eta\int\! \dd^2 x\,\sqrt{-\bar{g}}\,
\left(\,T^{(0)}_{\mu\nu}{T^{(0)}}^{\mu\nu}-{T^{(0)}}^2\right)\,. \label{TT-AdS2}
\end{align}
Note here that the deformation is measured by $\kappa\,\eta$\,. It is significant 
to see the signature of the deformation because it is sensitive to the physics of the deformed theory. 
Recall that both $\kappa$ and $\eta$ are positive. Hence the deformation (\ref{TT-AdS2}) 
corresponds to the negative sign in the convention of \cite{LST}. 
Then the deformed theory should have a UV cut-off (at least) in the flat-space limit, 
because the energy becomes complex in the UV region. 
Hence the above result would have an intimate 
connection with the cut-off AdS geometry \cite{Verlinde,Kraus} or the random boundary geometry \cite{Hirano}.

\medskip

On the other hand, a negative $\eta$ corresponds to a positive-sign $T\bar{T}$-deformation. 
Then the deformed theory does not have the UV cut-off. However, if $\eta$ is negative, then 
the potential of the dilaton is not bounded from below and the dilaton becomes unstable. 
This case may be interpreted as a quantum Liouville theory and then be related 
to the Little String Theory scenario proposed in \cite{LST}.

\subsection{The explicit solution of $\sigma$}

The remaining task is to derive a non-trivial solution to the equations of motion 
(\ref{eq:eom1f}) and (\ref{eq:eom2f}). For this purpose, 
let us start from considering some properties of the energy-momentum tensor.

\medskip

The energy momentum tensor $T^{(0)}_{\mu\nu}$ should satisfy the conservation law.
\begin{align}
\bnabla^\mu T^{(0)}_{\mu\nu}=0\,.
\end{align}
In the conformal gauge (\ref{conformal gauge}), the components of the conservation law 
are given by 
\begin{align}
\partial_- T^{(0)}_{++}=-\partial_+ T^{(0)}_{+-}-\frac{2}{x^+-x^-}T^{(0)}_{+-}\,, \qquad 
\partial_+ T^{(0)}_{--}=-\partial_- T^{(0)}_{+-}+\frac{2}{x^+-x^-}T^{(0)}_{+-}\,. 
\label{conservation}
\end{align}
The trace of the energy-momentum tensor ${T^{(0)}}_{+-}$ is not zero 
and gives rise to a no-trivial contribution.
Moreover, in comparison to the conformal matter case, the (++) component of 
the energy-momentum tensor ${T^{(0)}}_{++}$ is no longer a holomorphic function 
and it depends on $x^-$ as well. This is also the same for $T^{(0)}_{--}$\,.

\medskip

Thus, the equations in (\ref{eq:AP}) are rewritten as 
\begin{align}
{\rm e}^{2\bar{\omega}}\partial_+\left({\rm e}^{-2\bar{\omega}}\partial_+\sigma\right)=
& -\kappa\, T^{(0)}_{++}(x^+,\,x^-)\,,\no\\
{\rm e}^{2\bar{\omega}}\partial_-\left({\rm e}^{-2\bar{\omega}}\partial_-\sigma\right)=&
-\kappa\,T^{(0)}_{--}(x^+,\,x^-)\,,\no\\[-4pt]
\partial_{+}\partial_-\sigma+\frac{2}{(x^+-x^-)^2}\sigma=&\kappa\,T^{(0)}_{+-}(x^+,\,x^-)\,.\label{eq:AP with non-conformal}
\end{align}
It is useful to introduce a scalar function $M(x^+\,,x^-)$ as
\begin{align}
\sigma=\frac{M(x^+\,,x^-)}{x^+-x^-}\,.
\end{align}
The equations in (\ref{eq:AP with non-conformal}) are further rewritten as
\begin{align}
\partial_+^2M=& -\kappa\,(x^+-x^-)\, T^{(0)}_{++}(x^+,\,x^-)\,,\no\\
\partial_-^2M=& -\kappa\,(x^+-x^-)\, T^{(0)}_{--}(x^+,\,x^-)\,,\no\\
(x^+-x^-)\,\partial_{+}\partial_-M+\partial_+M-\partial_-M = 
&2\kappa\,(x^+-x^-)\,T^{(0)}_{+-}(x^+,\,x^-)\,. 
\label{eqM:AP with non-conformal}
\end{align}
The solution is given by
\begin{align}
M(x^+\,,x^-)=I_0(x^+\,,x^-)+{\cal I}^+(x^+,\,x^-)-{\cal I}^-(x^+,\,x^-)\,,\label{LiouvilleM}
\end{align}
where $I_0(x^+\,,x^-)$ is the sourceless solution given in (\ref{sourceless part}).
${\cal I}^+(x^+,\,x^-)$ and ${\cal I}^-(x^+,\,x^-)$ are defined as
\begin{align}
{\cal I}^+(x^+,\,x^-)&\equiv\frac{\kappa}{2}\int^{x^+}_{u^+}\dd s\,(s-x^+)(s-x^-)\,T^{(0)}_{++}(s,\,x^-)\,,\\
{\cal I}^-(x^+,\,x^-)&\equiv\frac{\kappa}{2}\int^{x^-}_{u^-}\dd s\,(s-x^+)(s-x^-)\,T^{(0)}_{--}(x^+,\,s)\,.
\end{align}
This solution resembles the one in the AP case (\ref{API}).
However, the energy-momentum tensor is not (anti-)holomorphic, 
hence be careful for calculating partial derivatives of ${\cal I}^\pm$\,.

\medskip

It would be instructive to demonstrate, for example, the calculation of 
the partial derivative of ${\cal I}^-$:
\begin{align}
&\partial_+{\cal I}^-(x^+,\,x^-)\no\\
&=\frac{\kappa}{2}\int^{x^-}_{u^-}\!\!\dd s\,\Bigl[-(s-x^-)T^{(0)}_{--}(x^+,\,s)+(s-x^+)(s-x^-)\partial_+T^{(0)}_{--}(x^+,\,s)\Bigr]\,\no\\
&=\frac{\kappa}{2}\int^{x^-}_{u^-}\!\!\dd s\,\Bigl[-(s-x^-)T^{(0)}_{--}(x^+,\,s)-(s-x^+)(s-x^-)\partial_sT^{(0)}_{+-}(x^+,\,s)-2(s-x^-)T^{(0)}_{+-}(x^+,\,s)\Bigr]\,\no\\
&=\frac{\kappa}{2}\int^{x^-}_{u^-}\!\!\dd s\,\Bigl[-(s-x^-)T^{(0)}_{--}(x^+,\,s)-(x^+-x^-)T^{(0)}_{+-}(x^+,\,s)\Bigr]\,.
\end{align}
From the second line to the third line, we have used the conservation law (\ref{conservation}) 
and also assumed that the boundary terms vanish.
Similarly, one can also evaluate the second-order derivative as follows: 
\begin{align}
&\partial_+^2{\cal I}^-(x^+,\,x^-)\no\\
&=\frac{\kappa}{2}\int^{x^-}_{u^-}\dd s\,\Bigl[-(s-x^-)\partial_+T^{(0)}_{--}(x^+,\,s)-T^{(0)}_{+-}(x^+,\,s)-(x^+-x^-)\partial_+T^{(0)}_{+-}(x^+,\,s)\Bigr]\,\no\\
&=\frac{\kappa}{2}\int^{x^-}_{u^-}\dd s\,\Bigl[-(s-x^-)\partial_sT^{(0)}_{+-}(x^+,\,s)-2\frac{(s-x^-)}{x^+-s}T^{(0)}_{+-}(s,\,x^-) \no \\ 
& \hspace*{3cm} 
-T^{(0)}_{+-}(x^+,\,s)-(x^+-x^-)\partial_+T^{(0)}_{+-}(x^+,\,s)\Bigr]\,\no\\
&=\frac{\kappa}{2}\int^{x^-}_{u^-}\dd s\,(x^+-x^-)\Bigl[-\frac{2}{x^+-s}T^{(0)}_{+-}(x^+,\,s)-\partial_+T^{(0)}_{+-}(x^+,\,s)\Bigr]\,\no\\
&=\frac{\kappa}{2}(x^+-x^-)T^{(0)}_{++}(x^+,\,x^-)\,,\no\\
&\partial_+\partial_-{\cal I}^-(x^+,\,x^-) =\frac{\kappa}{2}\int^{x^-}_{u^-}\dd s\,\Bigl[T^{(0)}_{--}(x^+,\,s)-T^{(0)}_{+-}(x^+,\,s)\Bigr]\,.
\end{align}
Thus, one can directly confirm the solution (\ref{LiouvilleM}) satisfies 
the equations in (\ref{eqM:AP with non-conformal}).

\section{Conclusion and discussion}

In this paper, we have revisited gravitational perturbations of the JT gravity and discussed 
the condition under which those can be seen as $T\bar{T}$-deformations. 
In addition to the known examples like the flat-space JT gravity and the AP model, 
as a novel example, we have studied the Liouville gravity with a negative cosmological constant.  
The conformal matter condition is necessary for the AP model but not for the Liouville gravity.  

\medskip

The Liouville gravity can also be seen as a Yang-Baxter deformation of the AP model. Then  
the parameter measuring the Yang-Baxter deformation is connected with 
the one of $T\bar{T}$-deformation, and also controls the behavior of the Liouville potential 
and the stability of the dilaton field. When the Liouville potential is bounded from below, 
the $T\bar{T}$-deformation is the negative sign $T\bar{T}$-deformation \cite{Verlinde}. 
The positive sign case may be related to a non-critical string approach 
\cite{Verlinde-NC, Tolley, Haruna}. It is also interesting to study 
a quantum aspect of our result by following \cite{quantum}. 

\medskip

It is an open problem to consider our result in the context of NAdS$_2$/NCFT$_1$ 
\cite{AP, Jensen, MSY, EMV}. As discussed in \cite{KOY}, the Yang-Baxter deformation brakes 
the $SL(2)$ symmetry and changes the UV behavior of the AdS$_2$ geometry. 
In particular, a singularity surface emerges at the middle of the bulk as a holographic screen and 
such a geometry would also be related to the cut-off AdS geometry proposed in \cite{Verlinde}. 
It is nice to study the boundary behavior of our non-local solution of the dilaton 
to figure out the boundary dual for the Liouville gravity.

\medskip

Finally, it is known that a single-trace $T\bar{T}$-deformation is related to a Yang-Baxter deformation 
in the context of AdS$_3$/CFT$_2$ \cite{Araujo, Borsato, Sfondrini1, Sfondrini2,Stijn}. 
It is interesting to understand a relation between this fact and our result via the dimensional 
reduction. It is also nice to consider a supersymmetric version of our analysis by following 
\cite{SUSY1,SUSY1-2,SUSY2,SUSY3}.

\subsection*{Acknowledgments}

It is grateful to thank R.~Tateo and A.~J.~Tolley for useful discussions 
and T.~Ishii and J.~Sakamoto for an earlier collaboration. 
The work of S.\,O.\ was supported by the Japan Society 
for the Promotion of Science (JSPS) No.\,18J21605. 
The work of K.\,Y.\ was supported by the Supporting Program for Interaction-based Initiative 
Team Studies (SPIRITS) from Kyoto University, and JSPS Grant-in-Aid for Scientific Research (B) 
No.\,18H01214. This work was also supported in part by the JSPS Japan-Russia Research Cooperative Program.

\appendix

\section*{Appendix}

\section{Derivation of the gravitationally dressed S-matrix}
\label{dressed S-matrix}

We shall derive here a gravitational dressing factor of the S-matrix. This was originally derived 
in \cite{Dubovsky} in the light-cone coordinates without the explicit solution of $\sigma$\,. 
It is instructive to reproduce the factor by using our exact solution of $\sigma$ with the Cartesian 
coordinates. 

\subsubsection*{Introducing the dynamical coordinates}

Let us first introduce the dynamical coordinates $X^\mu$ defined as
\begin{equation}
X^\mu \equiv -\frac{2}{\Lambda} \partial^\mu \phi = x^\mu + Y^\mu\,, 
\qquad Y^\mu \equiv -\frac{2}{\Lambda} \partial^\mu \sigma\,. 
\label{dynamX}
\end{equation}
The components of $Y^\mu$ are explicitly given by
\begin{align}
Y^t(t,x) =& \frac{2}{\Lambda} a_2+\kappa\, k \left[ x\, T^{(0)}_{tx}(t,x)+t\, T^{(0)}_{xx}(t,x)\right] \no\\
&+ \frac{\kappa}{\Lambda}(k\Lambda-2) \left( \int^{x}_{0} \dd x' \,T^{(0)}_{tx}(t,x') + \int^{t}_{t_2} \dd t'\, T^{(0)}_{xx}(t',0) \right), \label{Yt} \\
Y^x(t,x) =& -\frac{2}{\Lambda} a_3 -\kappa\, k \left[ x\, T^{(0)}_{tt}(t,x)+t\, T^{(0)}_{tx}(t,x)\right] \no\\
&- \frac{\kappa}{\Lambda}(k\Lambda-2) \left( \int^{t}_{0} \dd t' \,T^{(0)}_{tx}(t',x) + \int^{x}_{x_2} \dd x' \,T^{(0)}_{tt}(0,x') \right), \label{Yx}
\end{align}
where the indices have been lowered in the right-hand side. 
Then $Y^{\mu}$ satisfies
\begin{equation}
\partial_\mu Y^\nu = -\frac{2}{\Lambda} \partial_\mu \partial^\nu \sigma =\frac{2\kappa}{\Lambda}\left((1-k\Lambda)-\frac{k \Lambda}{2}\, x^\rho \partial_\rho \right)({T^{(0)}}_{\mu}^\nu-\delta_\mu^\nu {T^{(0)}})\,, 
\end{equation}
where we have used \eqref{eq:eom2}. 

\medskip 

In the standard manner, the conserved charge is given by 
\begin{equation}
P_\mu \equiv \int^{\infty}_{-\infty} \dd x \, T^{(0)}_{t\mu}(t,x)\,.
\end{equation}
The total energy $P_t$ and momentum $P_x$ are given by, respectively, 
\begin{equation}
P_t =  \int^{\infty}_{-\infty} \dd x \, T^{(0)}_{tt}(t,x)\,,  
\qquad
P_x =  \int^{\infty}_{-\infty} \dd x \, T^{(0)}_{tx}(t,x)\,.
\label{P_tx}
\end{equation}
The conservation law of the energy momentum tensor is given as
\begin{equation}
\partial^\mu T^{(0)}_{\mu\nu} = 0\,.
\label{tconserv}
\end{equation}
In the Cartesian coordinates, it is expressed as
\begin{equation}
\partial_t T^{(0)}_{tt} = \partial_x T^{(0)}_{tx}\,, \qquad
\partial_t T^{(0)}_{tx} = \partial_x T^{(0)}_{xx}\,.
\label{tcontx}
\end{equation}
Using the relations in \eqref{tcontx} and the invariance under the parity-transformation, 
\begin{equation}
T^{(0)}_{tx}(t,\infty) = T^{(0)}_{tx}(t,-\infty)\,, \qquad
T^{(0)}_{xx}(t,\infty) = T^{(0)}_{xx}(t,-\infty)\,, 
\end{equation}
the conservation of the charges $P_\mu$ is shown as follows; 
\begin{align}
\partial_t P_t = \int^{\infty}_{-\infty} \dd x \, \partial_x T^{(0)}_{tx}(t,x)=0\,, \qquad 
\partial_t P_x = \int^{\infty}_{-\infty} \dd x \, \partial_x T^{(0)}_{xx}(t,x)=0\,.
\end{align}

\medskip 

Note here that $Y^\mu$ still contains four arbitrary parameters $a_2$, $a_3$, $t_2$ and $x_2$\,. 
In order to fix the expression of $Y^{\mu}$ definitely, we need to impose some boundary conditions 
for $Y^{\mu}$\,. Then, as a result, \eqref{Yt} and \eqref{Yx} can be expressed 
in terms of the conserved charges $P_\mu$\,.

\medskip

Let us first impose a boundary condition for the energy momentum tensor as follows: 
\begin{equation}
x\,T^{(0)}_{\mu\nu}(t,x)\rightarrow 0\qquad (x\rightarrow \pm\infty)\,.\label{boundary conditon}
\end{equation}
By using the conservation of $T_{\mu\nu}^{(0)}$ in \eqref{tcontx}, one can obtain 
the following relations: 
\begin{align}
\int^{t}_{t_2} \dd t' \,T^{(0)}_{xx}(t',0) &= \int^{0}_{-\infty} \dd x' \,T^{(0)}_{tx}(t,x') - \int^{0}_{-\infty} \dd x' \,T^{(0)}_{tx}(t_2,x') + \int^{t}_{t_2} \dd t' \,T^{(0)}_{xx}(t',-\infty)\,, \\
\int^{t}_{0} \dd t' \,T^{(0)}_{tx}(t',x) &= \int^{x}_{-\infty} \dd x' \,T^{(0)}_{tt}(t,x') - \int^{x}_{-\infty} \dd x' \,T^{(0)}_{tt}(0,x') + \int^{t}_{0} \dd t' \,T^{(0)}_{tx}(t',-\infty)\,.
\end{align}
Then $Y^\mu$ can be rewritten as
\begin{align}
Y^t(t,x) =& \frac{2}{\Lambda} a_2 +\kappa\, k \left[ x\, T^{(0)}_{tx}(t,x)+t\, T^{(0)}_{xx}(t,x)\right]\no\\ 
&+\frac{\kappa}{\Lambda}(k\Lambda-2) \left( \int^{x}_{-\infty} \dd x' \,T^{(0)}_{tx}(t,x') - \int^{0}_{-\infty} \dd x' \,T^{(0)}_{tx}(t_2,x')\right), \label{Yt2} \\
Y^x(t,x) =& -\frac{2}{\Lambda} a_3-\kappa\, k \left[ x\, T^{(0)}_{tt}(t,x)+t\, T^{(0)}_{tx}(t,x)\right]  \no\\
&- \frac{\kappa}{\Lambda}(k\lambda-2) \left( \int^{x}_{-\infty} \dd x' \,T^{(0)}_{tt}(t,x') - \int^{x_2}_{-\infty} \dd x' \,T^{(0)}_{tt}(0,x') \right). \label{Yx2}
\end{align}
Now the unknown constants $a_2$ and $a_3$ are determined by using the boundary condition (\ref{boundary conditon}) as follows: 
\begin{align}
a_2 &=  \frac{\kappa}{2}(k\Lambda-2) \int^{0}_{-\infty} dx' \,T^{(0)}_{tx}(t_2,x') 
+ \frac{\Lambda}{2} Y^{t}_{(-)}\,, \\ 
a_3 &=  -\frac{\kappa}{2}(k\Lambda-2) \int^{x_2}_{-\infty} dx' \,T^{(0)}_{tt}(0,x') - \frac{\Lambda}{2} Y^{x}_{(-)}\,,
\label{a2a3}
\end{align}
where we have defined $Y^{\mu}_{(\pm)} \equiv Y^{\mu}|_{x \to \pm \infty}$\,. 
Using these expression of $a_{2,3}$, we find that 
\begin{align}
Y^t(t,x) &= Y^{t}_{(-)} +\kappa\, k \left[ x\, T^{(0)}_{tx}(t,x)+t\, T^{(0)}_{xx}(t,x)\right]+ \frac{ \kappa}{\Lambda}(k\Lambda-2)\, \int^{x}_{-\infty} \dd x' \, T^{(0)}_{tx}(t,x')\,, \label{Yt3} \\
Y^x(t,x) &= Y^{x}_{(-)}-\kappa\, k \left[ x\, T^{(0)}_{tt}(t,x)+t\, T^{(0)}_{tx}(t,x)\right] - \frac{ \kappa}{\Lambda}(k\Lambda-2)\, \int^{x}_{-\infty} \dd x' \, T^{(0)}_{tt}(t,x')\,. \label{Yx3}
\end{align}
Taking $x \to \infty$ and using \eqref{P_tx} leads to the following relations: 
\begin{equation}
Y^t_{(+)} - Y^t_{(-)} =  \frac{ \kappa}{\Lambda}(k\Lambda-2)\,P_x\,, \qquad
Y^x_{(+)} - Y^x_{(-)} = -\frac{ \kappa}{\Lambda}(k\Lambda-2)\,P_t\,.
\label{pmdiff}
\end{equation}
By employing a parity symmetric prescription, we obtain that 
\begin{equation}
Y^t_{(\pm)} = \mp \frac{ \kappa}{2\Lambda}(k\Lambda-2)\,P_x\,, \qquad 
Y^x_{(\pm)} = \pm \frac{ \kappa}{2\Lambda}(k\Lambda-2)\,P_t\,.
\end{equation}
It is useful to introduce a new quantity $\widetilde{P}_{\mu}$ defined as 
\begin{equation}
\widetilde{P}_{\mu} \equiv 2\,\int_{-\infty}^{x} \dd x \, T^{(0)}_{t\mu}(t,x)-P_{\mu}\,.
\end{equation}
In the spacial infinity region $x\rightarrow\pm\infty$, $\widetilde{P}_{\mu}$ becomes the conserved charge $\widetilde{P}_{\mu}\rightarrow\pm P_{\mu}$\,. 

\medskip 

Finally, the dynamical coordinates in \eqref{dynamX} are expressed in terms of $T_{\mu\nu}^{(0)}$ 
as follows: 
\begin{equation}
X^\mu = x^\mu -\kappa\,k\,({T^{(0)}}^{\mu}_\nu-\delta^\mu_\nu T^{(0)})\,x^\nu- \frac{\kappa}{2\Lambda}\left(k\,\Lambda-2\right) \epsilon^{\mu\nu} \widetilde{P}_\nu\,.
\label{dynamX2}
\end{equation}
Here $\epsilon^{\mu\nu}$ is an antisymmetric tensor normalized as $\epsilon^{tx}=-1$\,.
For simplicity, we will set $k=0$ in the following discussion.
Then, the metric fluctuation $h_{\mu\nu}$ and the quadratic action vanish while $\sigma$ does not.
The dynamical coordinates in (\ref{dynamX2}) are simplified as 
\begin{equation}
X^\mu = x^\mu + \frac{\kappa}{\Lambda} \epsilon^{\mu\nu} \widetilde{P}_\nu\,.
\label{dynamX3}
\end{equation}
This corresponds to the one obtained in \cite{Dubovsky}.

\subsubsection*{The gravitationally dressed S-matrix}

A significant implication of the dynamical coordinates in (\ref{dynamX2}) is 
the gravitationally dressed S-matrix \cite{Dubovsky}. 

\medskip 

Let us consider a scattering process in a scalar field theory. Here the detail of the interaction potential 
is not necessary. In the infinite past $t\rightarrow - \infty$\,, $N_{\rm in}$ particles are prepared and 
each of them has a momentum $p_{(i)}$\,. Then the asymptotic field (in-field) is given by
\begin{equation}
\psi=\int\!\! \frac{ \dd p}{\sqrt{2E}}\frac{1}{2\pi}\left[a^\dagger_{\rm in}(p)\,{\rm e}^{-ip_\mu x^\mu}+{\rm h.c.}\right]\,.
\end{equation}
It is known that a $T\bar{T}$-deformed QFT on the undeformed background 
is equivalent to the undeformed QFT with the dynamical coordinates 
\cite{Dubovsky,Cardy, Tateo2,Tateo3}.
This statement means that the deformation effect for the asymptotic state can be evaluated 
by replacing the original coordinates $x^\mu$ with the dynamical ones $X^\mu$\,. 

\medskip

As a result, a creation operator $a^\dagger_{\rm in}$ gets a extra-phase factor 
${\rm e}^{ip_\mu Y^\mu}$ and a dressed creation operator can be defined as  
\begin{equation}
A^\dagger_{\rm in}(p)\equiv a^\dagger_{\rm in}(p)\,{\rm e}^{ip_\mu Y^\mu}\,.
\end{equation}
By employing this dressed operator $A^\dagger_{\rm in}(p)$ 
(instead of $a^\dagger_{\rm in}(p)$)\,, the associated dressed in-state can be defined as  
\begin{align}
\Ket{\{p_{(i)}\},\,{\rm in}}_{\rm dressed}&\equiv\prod^{N_{\rm in}}_{i=1}A^\dagger_{\rm in}
(p_{(i)})\Ket{0}\,\no\\
&={\rm exp}
\left(i\,\sum^{N_{\rm in}}_{i=1}  {p_{(i)}}_\mu Y^\mu(x_{(i)})\right)
\Ket{ \{p_{(i)}\}, \, {\rm in}} \,.
\end{align}
In the infinite past, $Y^\mu(x_{(i)})$ can be evaluated as follows: 
\begin{align}
Y^\mu(x_{(i)})&=\frac{\kappa}{\Lambda} \epsilon^{\mu\nu} \left[2\,\int_{-\infty}^{x_{(i)}} \dd x' \, T^{(0)}_{t\nu}(t,x')-P_{\nu}\right]\no\\
&=\frac{\kappa}{\Lambda} \epsilon^{\mu\nu} \left[2\left(\frac{1}{2} {p_{(i)}}_\nu+\sum_{j<i} {p_{(j)}}_\nu\right)-\sum^{N_{in}}_{i=1} {p_{(i)}}_\nu\right]\no\\
&=\frac{\kappa}{\Lambda} \epsilon^{\mu\nu} \left( {p_{(i)}}_\nu+\sum_{j<i} {p_{(j)}}_\nu-\sum_{j>i} {p_{(j)}}_\nu\right)\,.
\end{align}
From the first line to the second line, we have assumed the mid-point prescription.
Finally, one can write the dressed state in terms of $p_{(i)}$.
\begin{align}
\Ket{\{p_{(i)}\},\,{\rm in}}_{\rm dressed}
&={\rm exp}
\left(2i\,\frac{\kappa}{\Lambda}\sum^{N_{\rm in}}_{i=1}\sum_{i<j}\epsilon^{\mu\nu} {p_{(i)}}_\mu {p_{(j)}}_\nu\right)
\Ket{ \{p_{(i)}\}, \, {\rm in}} \,.
\label{dressed in}
\end{align}
The phase factor in front of the original in-state is nothing but the gravitational dressing factor.
Similarly, the phase factor for the out-state can be evaluated and then 
the dressed S-matrix can be derived as shown in \cite{Dubovsky}.

\section{The vacuum solution in the deformed AP model}

Here, we introduce the general vacuum solution in the Yang-Baxter deformed AP model 
(\ref{deformedAP})\,.This is a short review of the result obtained in \cite{KOY}.

\medskip 

In the conformal gauge (\ref{conformal gauge}), the vacuum equations of motion 
(\ref{eq:veom1}) and (\ref{eq:veom2}) can be decomposed into a copy of two Liouville 
equations and a constraint condition as follows: 
\begin{align}
2\,\partial_+\partial_-\bar{\omega}_1+\frac{1}{L^2}{\rm e}^{2\bar{\omega}_1}&=0 
\label{Liouville1}\,,\\
2\,\partial_+\partial_-\bar{\omega}_2+\frac{1}{L^2}{\rm e}^{2\bar{\omega}_2}&=0 
\label{Liouville2}\,,\\
-{\rm e}^{2\bomega}\partial_\pm\left({\rm e}^{-2\bomega}\partial_\pm\bphi\right)&=0
\label{constraint}\,.
\end{align}
Here $\bar{\omega}_{1,2}$ are defined as
\begin{align}
\bar{\omega}_{1}\equiv\bomega+\eta\left(\frac{2}{L^2}\bphi-\Lambda\right)\,,\qquad\bar{\omega}_{2}\equiv\bomega-\eta\left(\frac{2}{L^2}\bphi-\Lambda\right)\,.
\end{align}
The general solution for each of the Liouville equations (\ref{Liouville1}) and (\ref{Liouville2}) 
are given by, respectively, 
\begin{align}
{\rm e}^{2\bar{\omega}_1}&=\frac{2\,L^2\,\partial_+X_1^+\partial_-X_1^-}{\left(X_1^+-X_1^-\right)^2}\,,\qquad{\rm e}^{2\bar{\omega}_2}=\frac{2\,L^2\,\partial_+X_2^+\partial_-X_2^-}{\left(X_2^+-X_2^-\right)^2}\label{generalX}\,. 
\end{align}
Here $X^+_{1,2}$ ($X^-_{1,2}$) are arbitrary holomorphic (anti-holomorphic) functions.
Moreover, $X^\pm_{1,2}$ must satisfy the condition (\ref{constraint})\,. 
Since (\ref{constraint}) can be rewritten by using the Schwarzian derivative ${\rm Sch}\{~,~\}$\,, 
\begin{align}
{\rm Sch}\{X_1^\pm,\,x^\pm\}-{\rm Sch}\{X_2^\pm,\,x^\pm\}=0\,, 
\end{align}
$X^\pm_{1}$ and $X^\pm_{2}$ are the same functions up to an $SL(2)$ transformation. 
Finally, the general vacuum solution in terms of $\bomega$ and $\bphi$ are given by $\bomega_{1,2}$ as
\begin{align}
{\rm e}^{2\bomega} ={\rm e}^{\bomega_1+\bomega_2}\,, \qquad 
\bphi =\frac{\Lambda\,L^2}{2}+\frac{L^2}{4\eta}(\bomega_1-\bomega_2)\,.
\end{align}

\subsubsection*{Example}

As an example, let us consider the following parametrization: 
\begin{eqnarray}
X^+_1(x^+) &=& \frac{ (1-\eta \beta)x^+ - 2 \eta \alpha }{-2 \eta \gamma x^+ 
+ (1+\eta \beta) } \,, 
\qquad X^-_1( x^-)  = x^-\,, \no \\
X^+_2(x^+) &=& \frac{ (1+\eta \beta)x^+ + 2\eta \alpha }{2\eta \gamma x^+ 
+ (1-\eta \beta) }\,,  
\qquad X^-_2(x^-) = x^-\,, 
\label{YB form}
\end{eqnarray}
where $\alpha$, $\beta$ and $\gamma$ are real constants. 
This vacuum solution describes the Yang-Baxter deformations of AdS$_2$ as follows: 
\begin{eqnarray}
{\rm e}^{2\bomega}&=&\frac{2L^2\,[1-\eta^2(\beta^2 +4 \alpha \gamma)] }{ 
(x^+-x^-)^2 - \eta^2 \left( 2\alpha +\beta (x^++x^-) -2\gamma x^+x^- \right)^2  }\,, \no\\\no\\
\bphi&=&\frac{\Lambda\,L^2}{2}+
\frac{L^2}{4\eta}{\text{log}}\left|\frac{ x^+-x^- + \eta\,( 2\alpha +\beta (x^++x^-) -2\gamma x^+x^-)   }{  
x^+-x^- - \eta\,( 2\alpha +\beta (x^++x^-) -2\gamma x^+x^-) }\right|\,. 
\end{eqnarray}
By taking the undeformed limit $\eta\rightarrow0$, this solution reduces to the general solution 
in the original AP model (\ref{APgeneral})\,.

\end{document}